\documentstyle[epsfig,psfig,draft]{mn}

\def\gs{\mathrel{\raise0.35ex\hbox{$\scriptstyle >$}\kern-0.6em
\lower0.40ex\hbox{{$\scriptstyle \sim$}}}}
\def\ls{\mathrel{\raise0.35ex\hbox{$\scriptstyle <$}\kern-0.6em
\lower0.40ex\hbox{{$\scriptstyle \sim$}}}}

\newcommand{\mum}{$\,\mu$m}

\newcommand{\etal}{et~al.\ }

\title[Adaptive optics near-IR imaging of NGC\,2992]
{Adaptive optics near-IR imaging of NGC\,2992 - unveiling core
structures related to radio figure-8 loops}

\author[Scott\,C.\ Chapman et al.]
       {Scott\,C.\ Chapman,$^{\! 1}$
        Simon\,L.\ Morris,$^{\! 2}$
        Almudena\ Alonso-Herrero,$^{\! 3}$
        Heino\ Falcke,$^{\! 4}$
        \vspace*{1mm}\\
        $^1$ University of British Columbia, Dept.\ of Physics \& Astronomy,
         Vancouver BC Canada, V6T 1Z4\\
        $^2$ Dominion Astrophysics Observatory, National Research Council of
                Canada, Victoria, B.C. V8X 4M6, Canada\\
        $^3$ Steward Observatory, the University of
                Arizona, Tucson, AZ 85721, USA\\
        $^4$ Department of Astronomy, University of Maryland College Park, MD
                20742-2421, USA;\\
                and Max-Planck-Institut f\"ur Radioastronomie, Auf dem H\"ugel
                69, D-53121 Bonn, Germany
        }

\date{Accepted ... ;
      Received ... ;
      in original form ...}

\pagerange{000--000}

\begin{document}

\maketitle

\begin{abstract}
We present near-IR adaptive optics, VLA radio
and HST optical imaging of the nearby Seyfert galaxy
NGC\,2992.
Spiral structure and an extension to the West are traced down to the
core region at the limiting resolution of our near-IR images.
A faint, diffuse loop of near-IR and radio emission is also observed to the
north, embedded within the prominent 2\arcsec
radio loop previously observed to the northwest.
Near-IR color maps, and CO narrowband imaging,
are then used to identify which regions may not be
purely reddened stellar populations.
Our new data provide evidence that the VLA radio-loop morphology in
the shape of a figure-8
represents two components superimposed:
1) outflow bubbles out of the plane of the disk, coincident with
the extended emission line region (EELR);
2) star formation along the spiral arm within the galaxy disk and through
the dust lane.
The near-IR continuum emission associated with the outflowing radio bubbles
suggest that the radio loops are driven by the active nucleus.
\end{abstract}

\begin{keywords}
   galaxies: active
-- galaxies: starburst
-- galaxies: formation
-- galaxies: individual: NGC2992
\end{keywords}

\def\gs{\mathrel{\raise0.35ex\hbox{$\scriptstyle >$}\kern-0.6em
\lower0.40ex\hbox{{$\scriptstyle \sim$}}}}
\def\ls{\mathrel{\raise0.35ex\hbox{$\scriptstyle <$}\kern-0.6em
\lower0.40ex\hbox{{$\scriptstyle \sim$}}}}

\section{Introduction}

NGC~2992 is an Sa galaxy seen almost edge-on with a nearby, likely interacting,
companion (NGC~2993).
It possesses an active Seyfert 1.9 nucleus.
A large and prominent dust lane runs through the center of the galaxy
roughly north to south,
splitting the nuclear region in two.
Ulvestad and Wilson (1984) found that the radio structure of the nucleus of
NGC~2992 has the shape of a ``figure-8'', with a maximum extent of about 
550\,pc, oriented out of the plane of the galactic disk
 (assumed distance 37.3 Mpc using the recession velocity relative
to the Local Group of 1864\,km\,s$^{-1}$ (Ward et al.~1980) 
and $H_0 = 50$ km~s$^{-1}$Mpc$^{-1}$, angular scale of
182\,pc / 1\arcsec\,).
Most of of the 6cm radio emission from the center of the galaxy arises in the
loops of the figure-8 rather than in the nucleus. 

There are several favored models for such figure-8 radio emission.
The loops could result from  expanding  gas bubbles  
which are seen preferentially as limb-brightened loops (Wehrle \& Morris 1988).
Such outflows may be associated with the AGN core. The continually
diminishing X-ray emission has been interpreted as a dying active nucleus 
(Bassani et al.~1998), possibly because fuel is no longer being channeled down to an
accretion disk region. 
However, this would not likely affect the appearance
of the surrounding region at our resolution of $\sim$20\,pc since even at 
outflow speeds close to $c$ the timescale is $\sim$ 50 years.

Alternatively starburst driven superwinds could result in expanding gas
bubbles leading to such emission (Heckman et al.~1990).
Extended X-ray and H$\alpha$ emission (Colbert et al.~1996, Colbert et al.~1998)
perpendicular to the plane of the galaxy
may be an indication for the superwind.
If the superwind were produced by an energetic burst
of supernovae in the past, the current radius of the loop would then imply
that the SNe explosion occurred over $2.35 \times 10^5$ years ago for a typical expansion velocity of 1000 km/s (Koo et al.~1992, Tenorio-Tagle et al.~1998).

A different  model proposes toroidal magnetic fields which result in a loop-like
emission of synchrotron emission (Wehrle \& Morris 1988). Here, strong differential rotation in the
galactic nuclear disk builds up the magnetic field having some radial 
component, until an instability occurs leading to an expanding magnetic arch.
The field configuration upon expansion away from the nucleus is then a pair
of loops.
The synchrotron emission results when particles are accelerated to
relativistic energies in the magnetic arches. 

We have observed NGC~2992 at high spatial resolution using the
Adaptive Optics Bonnette (AOB) on the CFHT. This is part of a larger  project
to map the cores
of nearby Seyfert galaxies in the near-IR
with adaptive optics
in order to study core morphologies.
Our new near-IR imagery of NGC~2992 unveils emission features
(dust obscured in optical wavelength HST imaging) which
may be related to the VLA radio emission.
This discovery allows a consistent model of the radio emission
to be formulated, incorporating the larger scale bi-polar morphology observed
in H$\alpha$ by Allen et al.~(1998).
In section 2 we outline the observations and reductions. In section 3 we present
the results of our multi-wavelength analysis. Finally in section 4, we discuss
the implications of the observed morphologies for various models.

\section{Observational Details}

\subsection{CFHT AOB imaging}
Observations were obtained using the CFHT in March, 1997 and 1998, with the Adaptive Optics Bonnette (AOB) (Rigaut et al.~1998) feeding the MONICA (1997) and KIR (1998) near-IR cameras (Nadeau et al.~1994). The AOB is based on the
curvature wavefront sensing concept (Roddier et al.~1991), and uses a
19 zone bimorph mirror to correct for wavefront distortions.
The MONICA detector is a Rockwell NICMOS3 array
with 256x256 pixels and 0.034"/pixel sampling.
The KIR camera replaced MONICA in early 1998, and uses
the Hawaii 1k x 1k pixel array with the same pixel sampling. 

The core of this galaxy is sufficiently bright as a guiding source to enable 
near diffraction limited resolution under favorable natural seeing conditions.
AO correction is much more efficient in the near-IR as a result of
the $\lambda^{1.2}$ dependence of the Fried parameter, $R_0$, which 
characterizes the atmospheric coherence length.
We obtained an H-band image with MONICA under good observing conditions
(seeing FWHM $<$ 0\farcs6 resulting in 0\farcs15 resolution.
The $J$, $H$, $K$- band and narrow-band $CO$ images obtained with KIR had worse
natural seeing ($>$0\farcs9) and the corrected PSF has
a FWHM of $\sim$0\farcs45 at $H$, $K$ and $CO$, and $\sim$0\farcs60 at $J$.
When forming colour maps, images are always smoothed to the worst resolution 
of the pair of images in question.

As the field size is small (9"x9" for MONICA, 36" x 36" for KIR), blank sky 
images were taken intermittently
between science frames. On-source images were taken in a mosaic of 4 positions,
alternately putting the galaxy core in each of the four quadrants of the
array to account for bad pixels and image uniformity.
Flux and PSF calibrations were performed using the UKIRT standard stars
fs13 and fs25. Flat-field images were taken on the dome with the lamps turned
on and off to account for the thermal glow of the telescope.
Data reduction was performed in the standard way for near-IR imaging:
{\it i)} bad pixel correction; {\it ii)}
sky  subtraction,  using a  median-averaged  sky estimate; {\it iii)} flat-field
correction;  {\it  iv)}   re-centering  and co-addition of  the  different
exposures   through cross-correlation  techniques.

The continuum image for the vibrational transition $CO$(2-0) absorption 
($\lambda_c$ = 2.296\mum\, $\Delta\lambda$ = 200\,\AA) was estimated by fitting 
a line through several regions in the $J$, $H$, and $K$ images with no
apparent dust-lane structure. All images were convolved to 0\farcs6 
resolution before estimating the CO\,continuum image. 
This extrapolation was then sampled with a Gaussian of
the same width as the $CO$ narrow filter centered at 2.296$\mu$m. This flux 
level was then used to normalize the $K$-band image for subtraction of the
$CO$ image, resulting in a [continuum - $CO$] index within the range typically
observed (see for example Davidge \& Courteau, 1999 for results on M81
with the same $CO$ filter used at CFHT).
The above result was also checked against the following procedure.
The outer regions of the image have typical early-spiral bulge colours 
(e.g. Glass \& Moorwood 1985) and are likely
to be relatively free of dust emission. 
By matching the $K$-band
image in flux to the $CO$ image near the outer edges of the field,
we obtain a zero-point in the [continuum - $CO$] image. 
These two procedures were found to agree within 15\%.

Obtaining an accurate measure of the resulting PSF of the image with AO is 
problematic, since the atmospheric conditions are continually changing, and 
a stellar image taken before or after the science exposure will likely not 
resemble the PSF  of the science frame. An accurate PSF must be temporally and 
spatially coincident with the actual region of interest in the field, since 
the PSF degrades away from the guiding source. Even when there is a star  
within the relatively small field of view of the near-IR cameras used with 
AOB, the PSF is frequently quite different from the core of the active galaxy 
(guiding source). Attempts have been made to model the off-axis PSF in these 
cases with some success (Hutchings et al.~1998). 

A technique has been developed 
to reconstruct a model PSF for the nuclear region of the galaxy  
using the adaptive optics modal control
loop information obtained during the actual observations (Veran et al.~1998a,1998b). 
For the core brightness as seen by the wavefront sensor (V=16.5) in the MONICA H-band image, simulations have shown that the reconstructed PSF should match the 
true PSF to an accuracy of approximately 10\% (Veran et al.~1998).
The factors degrading such a reconstruction are mostly the faintness and extension of the particular galaxy core, and superior results are obtained with 
brighter, point-like galaxy nuclei (or stars). 
The error in our reconstruction is  mostly in the Strehl ratio and outer 
artifacts with the FWHM being very close to that of the actual PSF. 

The image is then deconvolved to the point at which the LUCY algorithm 
(Lucy 1984) converges 
($\sim$25 iterations), using the model PSF as input. 
The LUCY deconvolution can create ringing artifacts around bright unresolved 
galaxy cores. However the core of NGC\,2992 is extended 
while many of the structures surrounding the core are point-like. This is ideal
for LUCY since the core region consists essentially of point sources superposed
over a varying background. 
The algorithm was found to provide a believable deconvolution, 
with no new features appearing
which were not in the original image at some level.
The main benefit is the reduction of sidelobes, and the seeing {\it pedestal} 
present with achieved Strehl ratios in the range 20-30\%.
The estimated 
final resolution in the image is 0\farcs12 corresponding to 4 pixels.

\subsection{Existing HST and UKIRT data}

An archive HST F606W filter image was obtained with the WFPC2 camera 
in 1994 as part of a snapshot survey
of nearby active galaxies (Malkan 1998). The pixel scale with the PC camera
is 0\farcs044 per pixel. In order to compare this directly with our AOB image, we
have interpolated and rebinned the image after rotating it to the proper 
orientation. 
The prescription to calibrate the HST F606W image to Johnson V-band
as outlined in Malkan et al.~(1998) is used for NGC\,2992. 

NGC\,2992 was also observed at the United Kingdom InfraRed Telescope (UKIRT) in the near-IR bands
 J, H and K. The plate scale is 0.29\arcsec /pixel. These data were 
originally presented by Alonso-Herrero et al.~(1998).

\subsection{VLA radio maps}
A VLA image at 6\,cm (5\,GHz) was obtained in 1987 by A. Wehrle and is reproduced here with her permission (figure 1, right panel).

An image at 8.4GHz (3.57cm) was also obtained from the VLA archive
(courtesy of H. Falcke) in 
C-configuration which resolves more clearly into knot-like structures
(figure 1, left panel).

%
%
\begin{figure}
\begin{center}
\epsfig{file=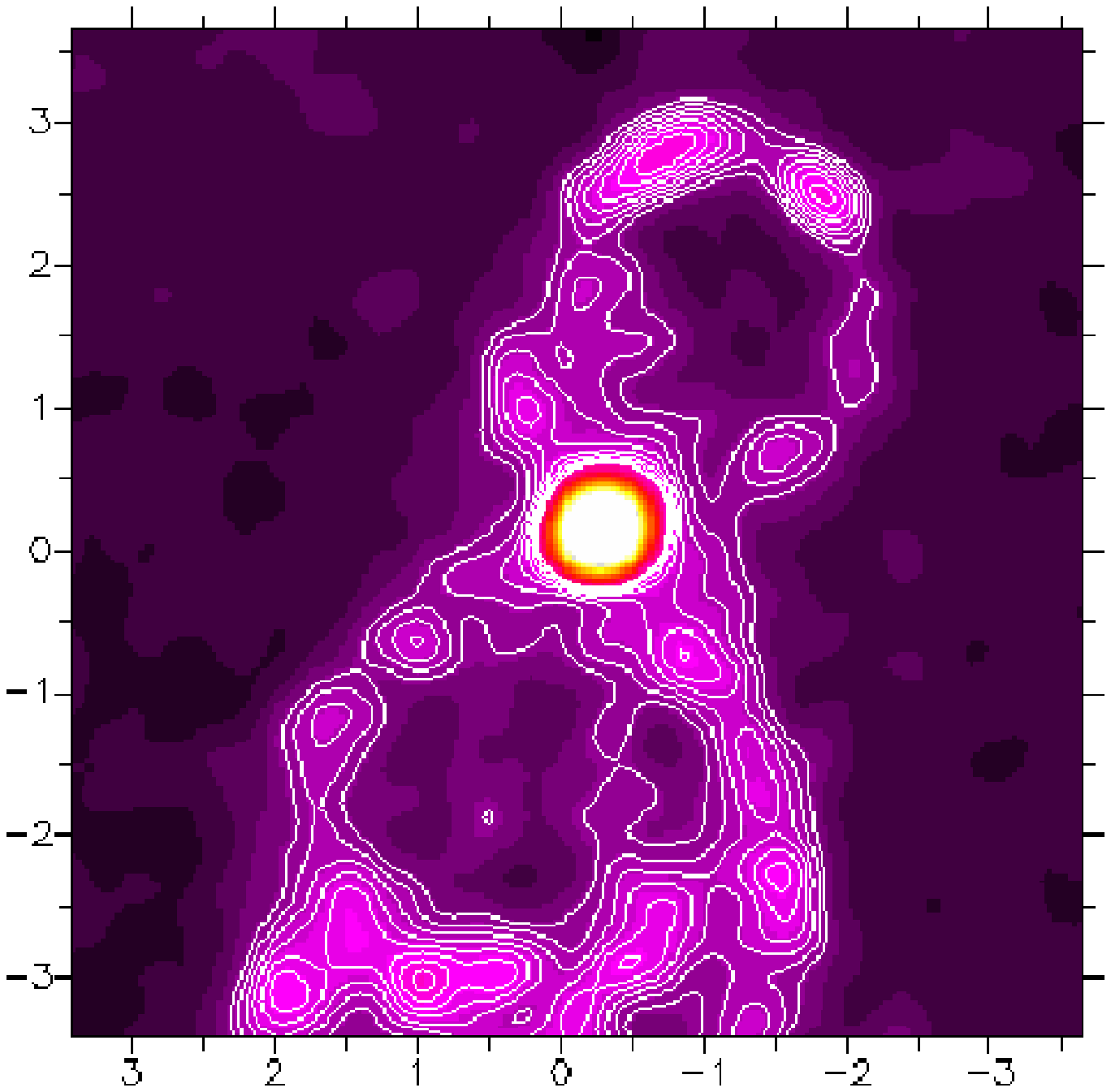, width=7.25cm, angle=0} 
\epsfig{file=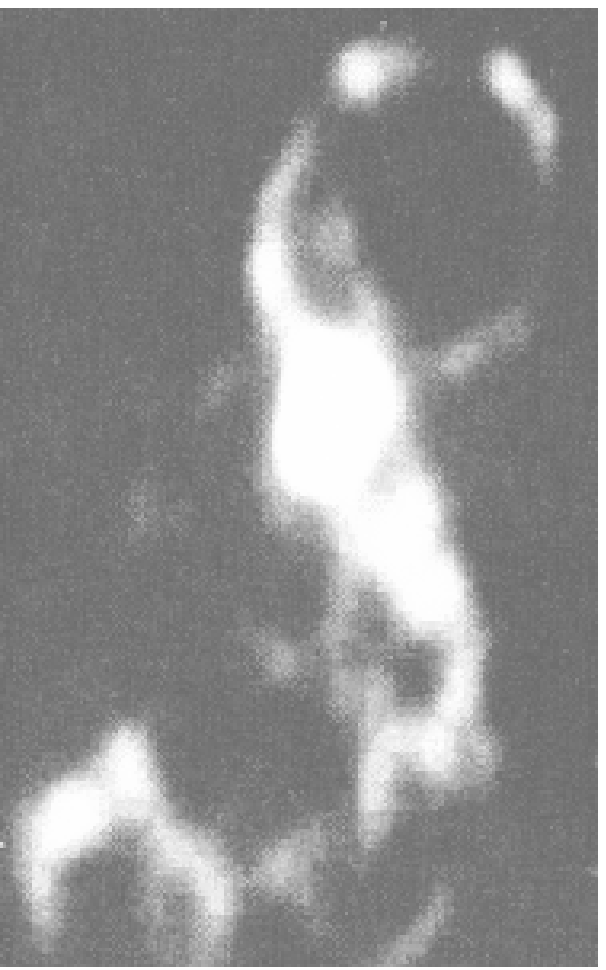, height=5.55cm, angle=0} 
\end{center}
\caption{Top: 8.4\,GHz image with contours overlaid (Peak flux: 7.0 mJy - 
contour levels from 0.14\,mJy to 0.54\,mJy, with intervals 0.045\,mJy). 
Bottom: reproduction of
the 5\,GHz image (Wehrle and Morris 1985) (Peak flux: 7.2 mJy).
The peak flux at 1.5\,GHz is 27.8 mJy (Ulvestad \& Wilson, 1984)}
\label{n2radioraw}
\end{figure}

%

\section{Results}

\begin{figure}
\begin{center}
\epsfig{file=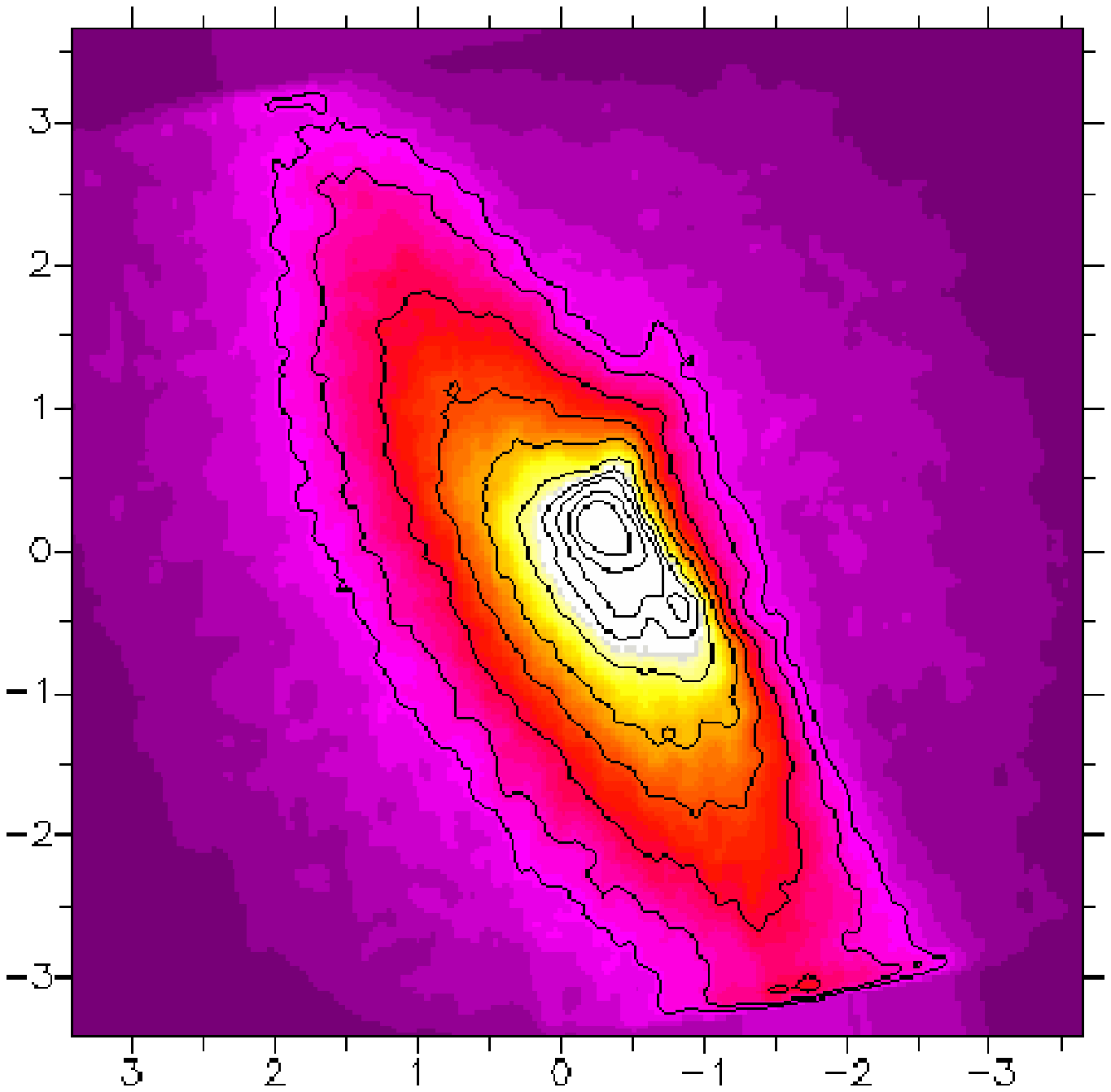, width=8.25cm, angle=0}
\epsfig{file=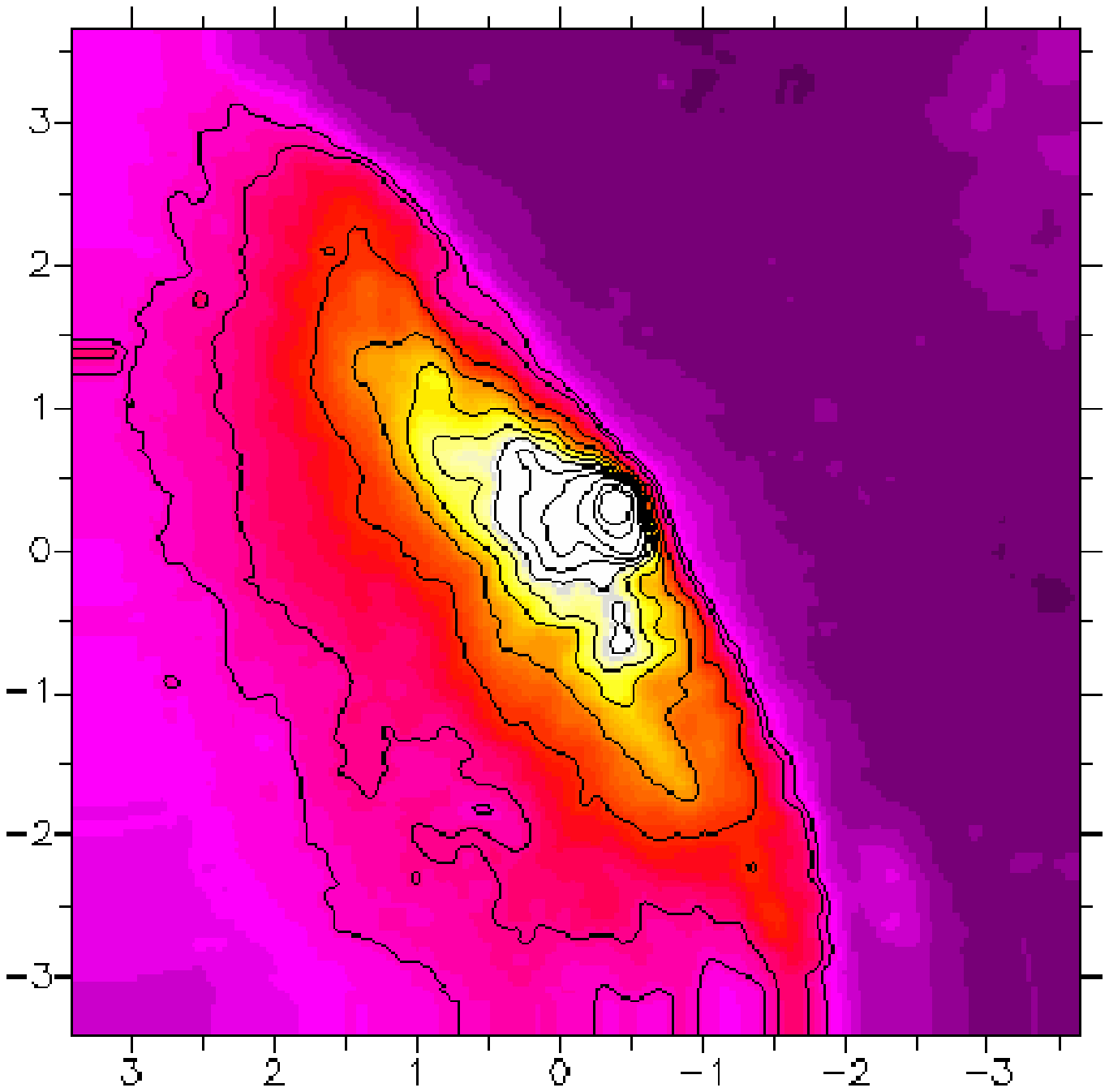, width=8.25cm, angle=0}
\end{center}
\caption{Upper panel: CFHT/AOB H-band image deconvolved with the LUCY algorithm. 
Lower panel: HST F606W (R+V band) image - subsequently called R-band. 
Contour levels are 1,3,5,8,11,15,20,30,50,80 percent of peak 
(H=14.1, F606W=17.7).
}
\label{raw}
\end{figure}

The morphology in the central region of NGC\,2992 is a complicated 
superposition of various
components, appearing most prominently in different wavelength regimes. 
An irregular-shaped patchy region dominates much of the western portion
of the shorter wavelength images (especially $R$-band -- see figure
\ref{raw} and subsequent images).
The fact that this region appears clearly as a deficit in the R-band image,
and takes on a
patchy morphology is strong evidence for dust obscuration as the source of the
colour gradient. The  region is therefore  most pronounced in long
baseline colour maps, such as R-H, with
longer wavelengths being least affected by dust.
Larger scale images of the galaxy reveal this feature as a dust-lane
structure bisecting the core region (see Wehrle \& Morris 1988).
In Table 1, we list the various structural components in the
core along with the figure which best highlights the feature.

%
%
\begin{table*}
\begin{center}
\caption{\hfil The structural components of NGC\,2992 \hfil}
\begin{tabular}{lccccl}
\noalign{\medskip}
\hline\hline
\noalign{\smallskip}
Feature & property & flux & size & best figure & location \\
\hline
\noalign{\medskip}
resolved core & SF+AGN & H=13 & 0.3\arcsec & All & \\
knots near core & SF & H=16 & unresolved & raw, smooth (4.2,4.5)&S of nucleus
 \\
\noalign{\medskip}
radio loops & & & 2\arcsec & 8.4\&5\,GHz (4.1) & NW,SE  \\
near-IR extension & nonstellar colours & H=15.3 & 1\arcsec & model subtract (4)  & NW  \\
\noalign{\smallskip}
inner loop & SF & & 1\arcsec & R-H (4.3) & N  \\
diffuse radio & SF & & 1\arcsec & radio (4.1) & N  \\
\noalign{\smallskip}
spiral arms & knots & & 3+\arcsec & model\& smooth (4.4,4.5) & NE-SW  \\
assoc. radio? & & & 3+\arcsec & radio (4.1) & NE-SW  \\
\hline
\end{tabular}
\end{center}
\noindent
\label{structures}
\end{table*}

%
%

%
%
\begin{figure}
\begin{center}
\epsfig{file=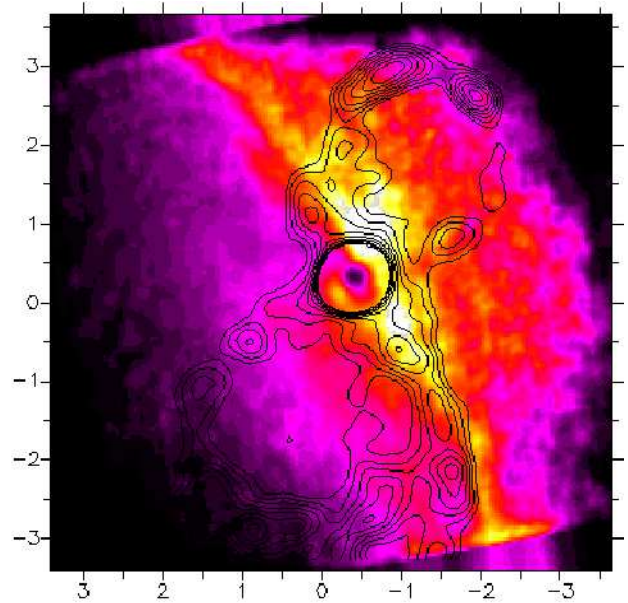, width=8.25cm, angle=0}
\end{center}
\caption{R-H colour map with 8.4~GHz contours overlayed. The radio map core 
centroid was registered to the near-IR core.
Brightest colours indicate the reddest regions with a maximum of R-H=4.8, 
where the nucleus R-H=3.6.
}
\label{colour}
\end{figure}

%
\begin{figure}
\begin{center}
\vskip 1cm	
\epsfig{file=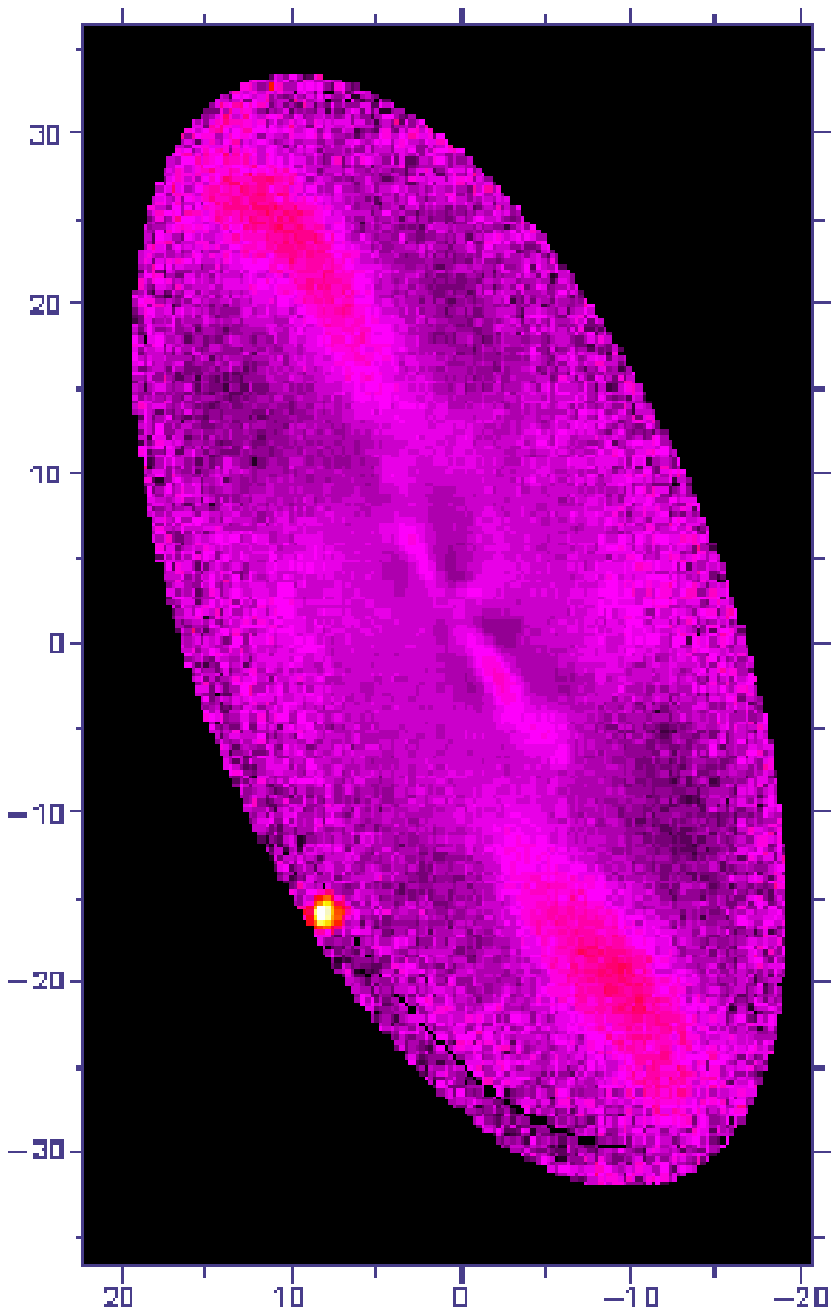, width=7.75cm, height=11cm, angle=0}
\epsfig{file=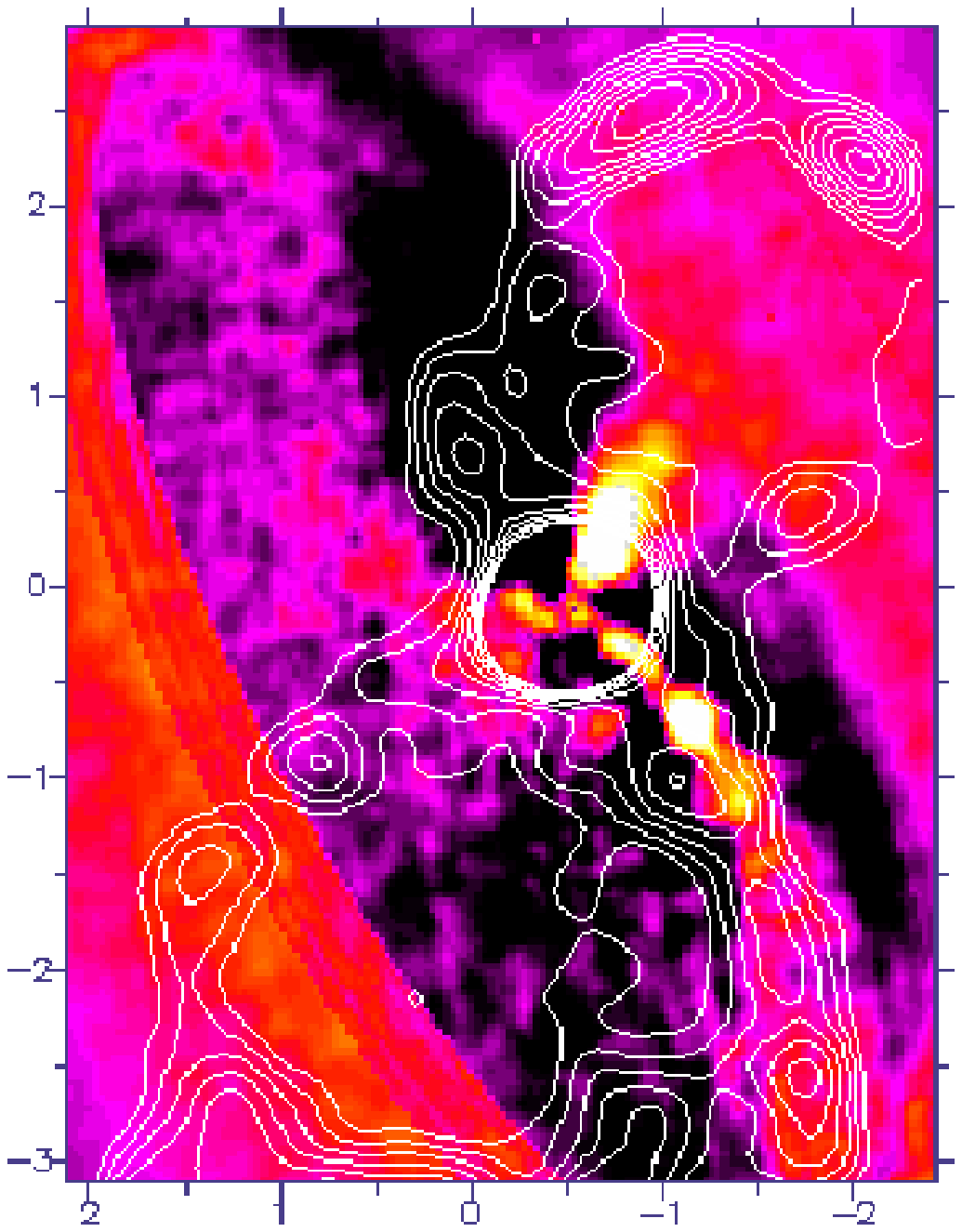, width=7.75cm, angle=0}
\end{center}
\caption{a) above: large-scale H-band UKIRT image with 0\farcs29 pixels, 
elliptical isophote model subtracted. 
b) below: AOB H-band isophotal model-subtracted image of the central 6\arcsec\
$\times$ 4\arcsec\ of NGC\,2992 with 8.4~GHz contours overlaid. }
\label{sub}
\end{figure}

%
%
\begin{figure}
\begin{center}
\epsfig{file=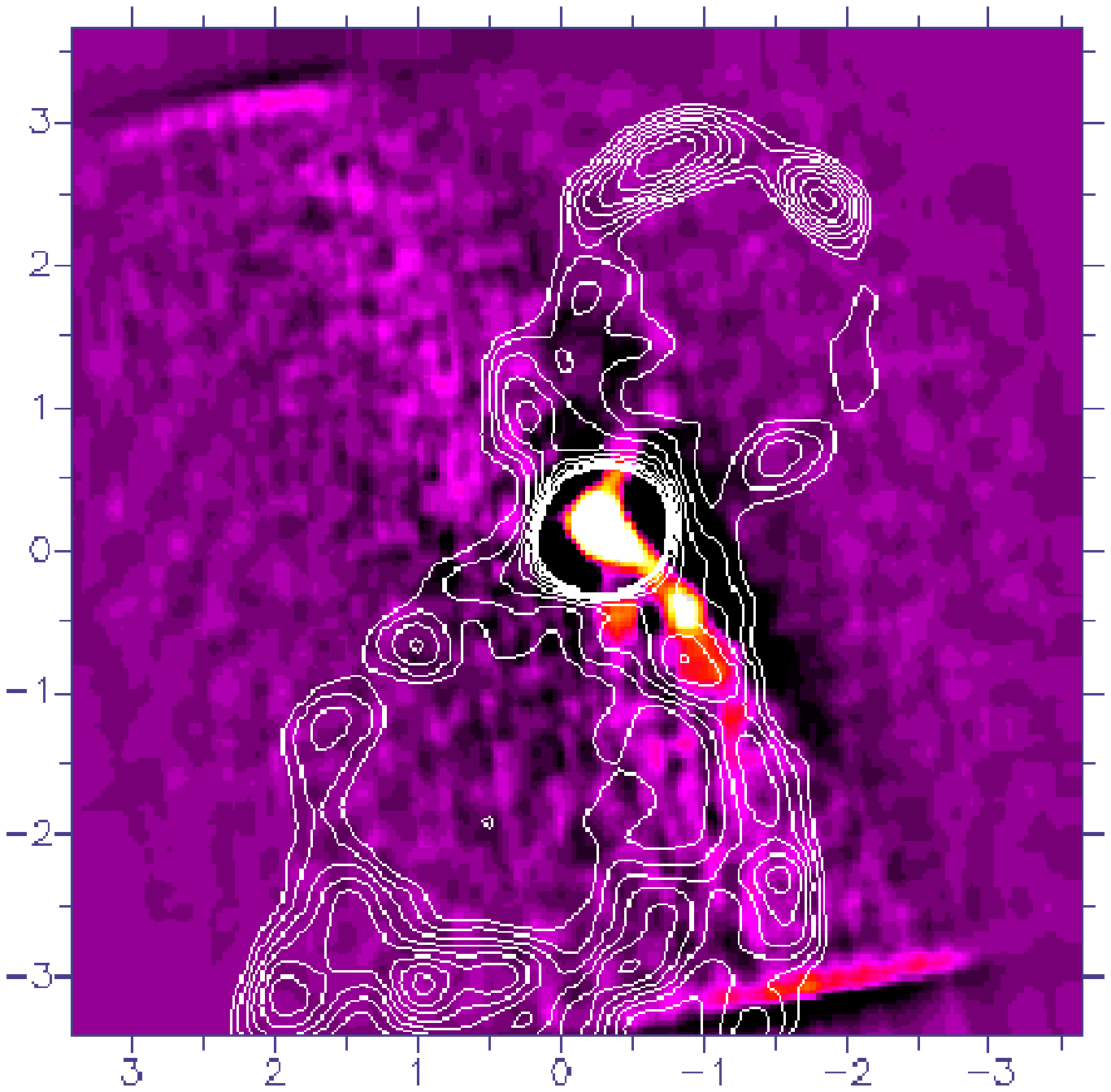, width=8.25cm, angle=0}
\epsfig{file=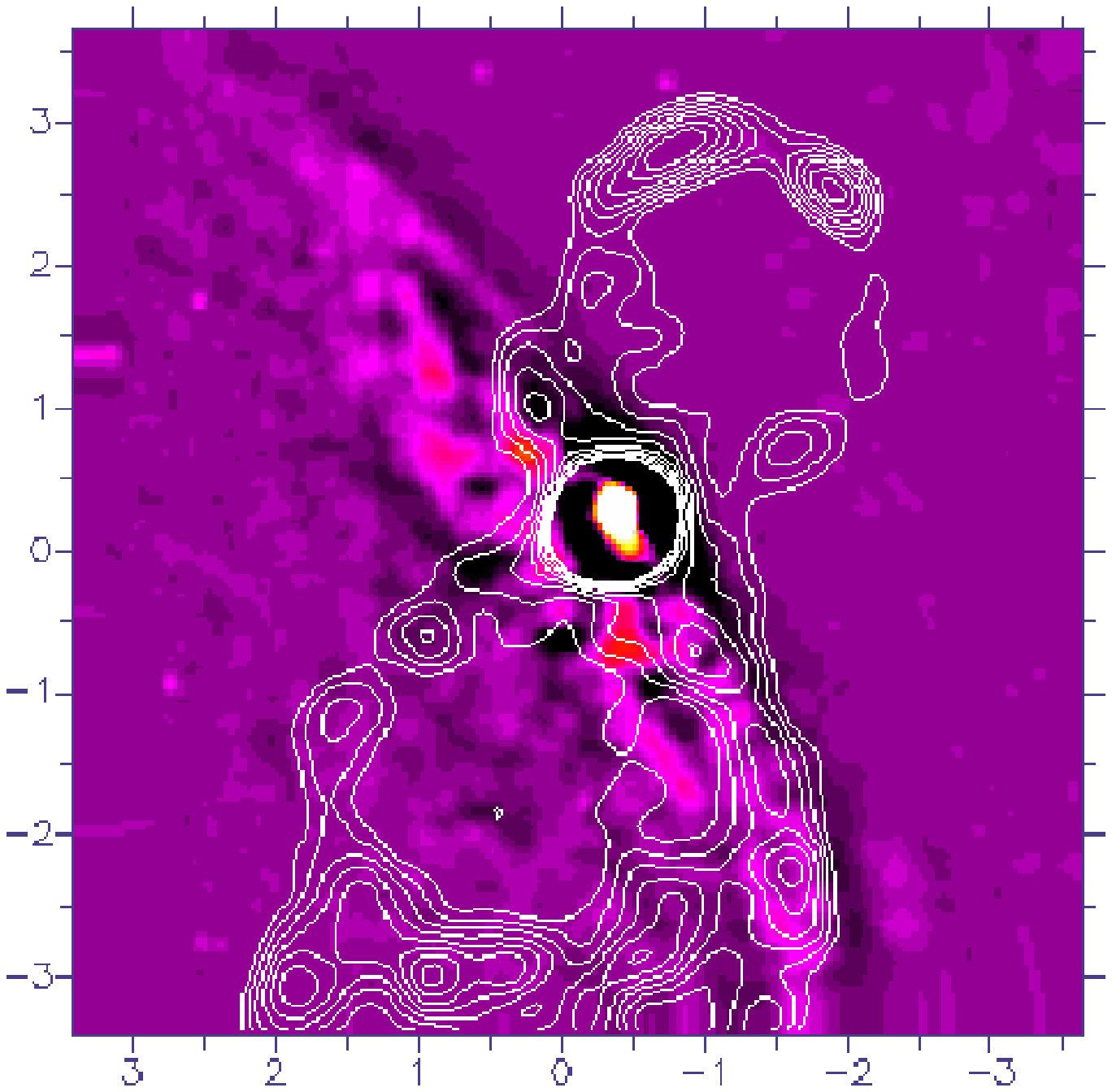, width=8.25cm, angle=0}
\end{center}
\caption{N2992 at H (above) and R-band (below), where a median filtered image
has been subtracted to remove the contribution of the low order galaxy
component. The 8.4~GHz radio contours have been overlayed.}
\label{filter}
\end{figure}

\subsection{The Core Region}
In figure \ref{raw}, 
we present the near diffraction limited H-band image of the central 7\arcsec\ of
NGC\,2992. The HST F606W (covering V+R band - hereafter called R) image is also depicted.
The galaxy core is resolved in both the H and R filters, with
a FWHM $\sim$0\farcs26 corresponding to 47.2\,pc.
For the mediocre correction in the KIR J and K images
the core is essentially unresolved with FWHM 0\farcs45 ($K$) and 0\farcs6 
($J$).  
It is important to understand what percentage of the nuclear light may
be emitted by an unresolved (AGN) component. 
We estimate the contribution to the extended galaxy core from the unresolved 
H-band component.  Our model-constructed PSF (see section 2 and 
Veran et al.~1998a) is a reasonable estimate of the true PSF for the 
diffraction limited H-band image. We first produce a surface
brightness profile of the galaxy using the IRAF task $ELLIPSE$, 
and another one for the PSF.  The PSF is then
scaled to match the peak of the galaxy brightness. Both
profiles are integrated out to a radial distance of 1\arcsec\ with the result
that the an unresolved component contributes at most 38\% at H in a 0.2\arcsec\ 
diameter aperture, and 17\% in a 0.5\arcsec\ aperture.


In Table 2 we present aperture photometry of the core region.
The V-H colour of the nucleus (V-H = 3.4) is typical for a reddened stellar
population (Glass \& Moorwood 1985).
Note that the deconvolved magnitudes agree with the
raw data at an aperture radius of 3\arcsec\,. This would be as expected given
that the deconvolution process restores flux to the central spike
from the surrounding pedestal.
This is encouraging despite the claims that deconvolved images
may not be appropriate for photometry since the non-linear algorithm does
not necessarily conserve flux (Magain et al.~1998).
The disagreement with Alonso-Herrero et al.~(1998) for the H-band apertures
less than 3\arcsec\ diameter
is likely a result of the improved resolution of our image coupled with our
absolute errors of 4\% for the RMS of calibration star measurements
throughout the observing run.

%
%
\begin{table}
\begin{center}
\caption{\hfil Aperture magnitudes for the AOB H-band image, and
HST F606W filter image corrected to V-band.\hfil}
\begin{tabular}{lcll}
\noalign{\medskip}
\noalign{\smallskip}
Aperture& H(deconvolved) &  H(raw) & V(HST) \cr
Diameter& & &  \\
\hline
\noalign{\medskip}
0\farcs5   &  14.08   &   14.50   & 17.46  \cr
1\farcs0    &  12.84   &   13.16   & 16.71  \cr
1\farcs5   &  12.29   &   12.50$^a$ & 16.22 \cr
2\farcs0    &  11.96   &   12.08   & 16.17  \cr
3\farcs0    &  11.56   &   11.57$^b$ & 16.14 \cr
\end{tabular}
\end{center}
\noindent
$^a$ H = 12.57 from Alonso-Herrero et al.~1998\\
$^b$ magnitudes agree past the 3rd Airy ring
\end{table}

The radio core spectrum is flat to within 5\% measured from our 8.4\,GHz VLA 
image, in conjunction with archive 5\,GHz and 1.5\,GHz VLA measurements 
(mean of $\nu F_\nu = 3.4\times10^{-16} ergs\,s^{-1}$)   
Wehrle \& Morris 1988, Ulvestad et al.~1984. This 
is indicative of synchrotron emission arising in the AGN core dominating over 
any possible synchrotron emission associated with supernovae at these 
wavelengths (Robson 1996).  

Examination of the cores in both optical and near-IR images (figures 2,
4 and 5)
reveals a bright point-like knot to the south $\sim$110\,pc
(0.6\arcsec) from the nucleus of the galaxy. A second knot  to the southwest
appears in the near-IR only.
An [R-H] colour map is formed by first convolving
the HST image with a Gaussian of width 0\farcs12 to match the AOB resolution.
The colour map (figure 3) reveals the southwestern
knot is almost one magnitude redder in R-H than the southern knot, likely as
a result of dust. 
The SW knot is completely obscured in the HST image.
We find the colours in these regions (V-H = 3.6 to 4.2) are consistent with
reddened stellar populations.

We perform aperture photometry on these knots in H-band 
find apparent magnitudes between 16.0 and 16.5 at H depending on aperture size 
used (the aperture measurements have relatively large error due to their
proximity and overlap with the central AGN
core). At the distance of NGC\,2992, the absolute magnitudes lie between 
$-15.3$ and $-14.8$ which are comparable to the luminosities 
of stellar clusters 
detected in the interacting galaxy Arp299, which have an average H-band absolute
magnitude of $H=-15$ (Lai et al.~1998, Alonso-Herrero et al.~1998).              
Arp220 (Scoville et al.~1998) also has compact stellar clusters of similar luminosities at 1.6$\mu$m (H-band). In the latter work, merger remnants have
 been suggested as an explanation for such bright knots near the galaxy core ($>10$\% of the AGN peak for NGC\,2992).                 

\subsection{Spiral Structure}

The H- and R-band images (figure 2) show elongated
isophotes to the southwest along the galaxy disk, and extending east from
the nucleus.
In the R-band case 
the galaxy morphology is much more distorted due to the effects of dust.
These extensions and distortions in the isophotes
suggest structure underlying the elliptical symmetry of the disk within the central regions of galaxy.
We subtract a model image of the galaxy for both a large-scale H-band image
(60\arcsec $\times$ 60\arcsec)
and the AOB H-band image. 
The model is built from either elliptical fitted isophotes (figure 4), 
or a running median filter image of FWHM twice the resolution (figure 5). 
The elliptical isophote model has the advantage of highlighting structures
deviating substantially from elliptical symmetry, while the median filtering
tends to bring out fainter point-like structures.
The profile in ellipticity ($E$) and position angle ($PA$) of the isophotal 
model shows that this region has a substantial twist in both $E$ and $PA$.
In figure 4 (upper panel), the model subtracted large-scale image displays 
what appears to be a spiral structure along the
disk, as well as a $\sim$3\arcsec\ extension  to the west, also noted in 
Alonso-Herrero et al.~(1998). 
Our new high resolution images show that both the spiral structure and
western extension can be traced down to the very core. This latter extension
will be taken up in the next section.

The image of the larger-scale galaxy shows a break in the spiral structures
at $\sim$7\arcsec\ radius. This may indicate nested structures which are kinematically distinct, as in the multi-level spiral arm structure observed
with adaptive optics in NGC\,5248 (Laine et al.~1998). 
Indeed, the knotty morphology of the spiral structures seen in the AOB and
HST images of NGC\,2992 are suggestive of star formation in spiral arms.
The high inclination of NGC\,2992 makes deprojection unreliable and it is difficult to discern how these larger- and smaller-scale spiral structures relate.
By the same measure, it is also difficult to know whether the isophotal twists 
in the core may be due to a bar, a triaxial bulge seen in projection, or
simply the effect of the spiral arms (Friedli et al.~1996).

In figures 4 and 5, the 8.4\,GHz radio contours are overlayed on the
model-subtracted images.
There is clearly some radio emission coincident with the
southern spiral arm, which breaks up into a similar knotty morphology to
the H-band model-subtracted image. 
Note especially the strong near-IR peak at the
southern tip of the spiral arm has an associated peak in the radio, 
$\sim$3\arcsec\ south and 2\arcsec\ west of the nucleus.
We find the radio spectrum to be steeper along the spiral arm to the south
than in the core, possibly indicative of star formation.

The prominent southern knots near the core, outlined in the previous section,
clearly lie along the southern spiral arm.
By comparing the brightest knots within the spiral arms in the optical and 
near-IR images, it appears that the nuclei of the galaxy in H- and R-bands,
determined by centroiding on the cores, do not line up precisely.
The knots in the southern spiral arm, revealed in the model subtracted images,
have similar
geometric configurations in the two filters (H and R), but when these
are superposed, there is a an (x,y) offset of 
(0\farcs3, 0\farcs15) between the
nuclei corresponding to a distance of  $\sim$70\,pc. 
A true offset is possible due to dust obscuration and/or extended unresolved 
starbursts (see for instance NGC\,1068 (Alloin et al.~1998 for a similar case).
Given the much more distorted  nuclear region in the HST R-band image compared 
to the H-band, such a scenario is plausible.  
However, the spiral arm morphology may be slightly different in the two 
wavelengths, due to the effects of both dust and unresolved, blended structures.
The difficulties in matching structures with slightly different morphologies
in the outer galaxy leads us to register the images by centroiding on the 
galaxy nucleus.          

\subsection{Figure-8 Loops}

There is little sign of optical or near-IR counterparts to the
radio loops out past the disk of the galaxy, even at K-band  where the
ability to see through the dust lane is greatest.
However, as noted above, there is an extended feature to the northwest,
observed as extended contours in the near-IR (figure 2) and as a prominent
feature in the model subtracted near-IR images (figure 4), extending 
$\sim$1.5\arcsec\ before becoming
more diffuse and mixing with the stellar emission on the western side of the 
dust-lane.
In the optical HST image (figures 2b, 5b), there is no sign of this extended 
feature, likely due to the dust lane obscuration.
This extension aligns with the mouth of the 
northern radio loop and appears to continue outwards into the loop,
which is verified in the coarser resolution UKIRT image where the S/N
allows tracing the feature out further.
This indicates that the source of the radio loop may be connected to 
this feature.

Figure \ref{colour} shows that the extended feature is the reddest
region in the central 7\farcs0 of the galaxy with an R-H colour of 4.5.
The lower resolution KIR images in J,H,K show that the colours in this
region clearly stand out from the surrounding disk colours, possibly as 
non-stellar (in a 1\arcsec\ aperture, J-H=0.5, H-K=1.0).
These colours may be related to a highly reddened ($A_V > 5$) burst of stars, 
possibly with a nebular component. However they are also consistent with a 
reddened continuum power-law emission (Glass \& Moorwood 1985).

Artifacts associated with the AO correction have been shown to produce
extensions to the PSF when guiding on extended objects such as as Seyfert
nuclei (Chapman et al.~1998). Although caution must  therefore be taken in 
associating such a structure with a physical interpretation, we have compared
our AOB H-band image with an HST NICMOS K-band image, and found the same
extended isophotes to the west of the core.

\subsection{Diffuse Inner Loop}

The R-H map (figure 3) shows extended red emission which takes on a loop-like
morphology extending north of the nucleus from the 1\arcsec\ to the 2\arcsec\ 
declination offsets, an enhancement in H rather than a deficit in R. 
It does not appear to be associated with the spiral arm further to the east.
In the median filtered H-band image (figure 5), it is also possible to discern 
knotty features in a loop-like morphology, however the emission lies deep 
within the region taken to be a dust-lane based on the obscuration of the
HST optical imagery. No counterpart is seen
in the model subtracted R-band (figure 5).

The 8.4\,GHz radio contours superposed over the above figures (3 \& 5)
reveal a similar loop-like diffuse emission embedded within the larger, well defined radio loop. 
The spectrum (from 5GHz/8.4GHz) is steeper here than in the core,
consistent with star forming regions.
Buried star formation regions have been identified in the dust lane crossing
the nuclear region of Centaurus A (Schreier et al.~1996,1998), using HST-NICMOS images in H-band.
The J,H,K colours in this region of NGC\,2992 however are difficult to 
interpret with the dust absorption gradient across the dust lane.
It is not clear that the structure has a true {\it loop} morphology, and may
be simply a result of the way the dust lane cuts through the core region.
However, the coincidence of the radio emission suggests that the near-IR excess may represent more than the artifact of dust absorption.

\subsection{The CO map}
%
%
\begin{figure}
\epsfig{file=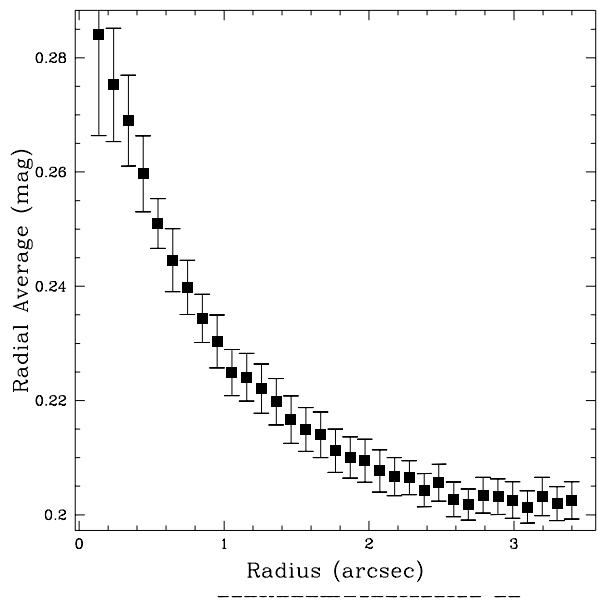, width=8.25cm, angle=0}
\caption{The CO index profile, (ctm - CO), near the center of NGC2992,
where distances are measured with respect to the galaxy nucleus.
The data points are azimuthal averages, and the error bars show the systematic errors introduced by uncertainties in the background.} 
\label{n2co}
\end{figure}

The radial distribution of the CO index, (continuum~--~CO), 
is shown in figure 6, 
where the plotted values are azimuthal averages in 0.1\arcsec\ bins.
Although the CO(2-0) filter used corresponds to the standard photometric index 
(2.296\mum\ center, 200\,\AA width) (Doyon et al.~1994), the subtracted 
continuum was extrapolated from the broad band colours (see section 2). 
There is then likely a systematic offset in absolute photometry, and we
use caution in interpreting the CO magnitudes. 

When $r <$ 1.5\arcsec\ the CO index strengthens
implying much more CO absorption within the central 3\arcsec\,.
The CO index depends on both the metallicity and the age of the stellar 
population (giants versus supergiants), and has been found to be difficult 
to model.  If we assume that the metallicity is 
constant within the center, then the variation may be an age effect. In 
general, the CO index increases as the starburst ages (Vanzi, Alonso-Herrero \& Rieke 1998, Doyon et al.~1994).
The NGC\,2992 CO profile suggests that a somewhat younger population is present in a ring around the galaxy center, while the stellar population in the very core would be slightly older than the surroundings.
Given the rather poor resolution in these images (0.45\arcsec ), we are unable
to discern at what level the stellar population contributes within the
resolved 50pc core, evident in the HST R- and MONICA H-band images.
However, it is clear that hot dust ($\sim$ 1000 K) emission does not contribute significantly to the core region.
The power-law tail of a strong hot dust continuum contribution at
K-band would swamp the CO absorption signature, leading to a weakening 
CO index (less CO absorption).
At larger distances the CO index shows a slight radial gradient, but this is
not significant given the uncertainties in the background.

\subsection{Characterizing the Extinction}

%
%
\begin{figure}
\psfig{file=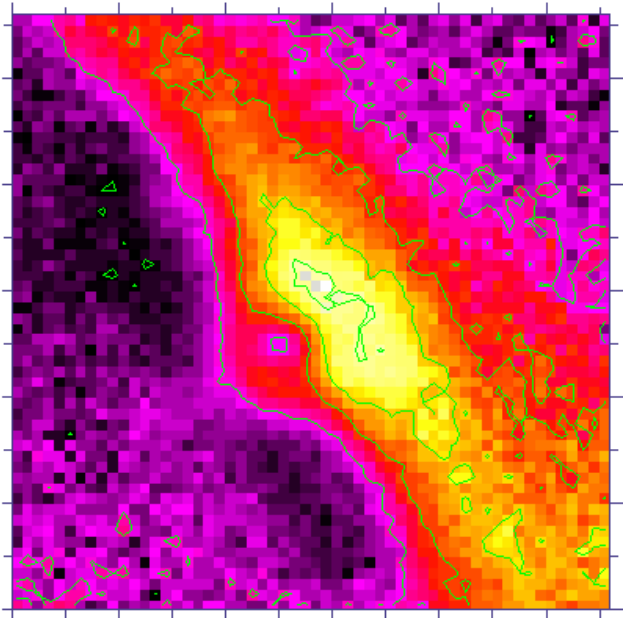,width=8.25cm,angle=0} 
\caption{NGC\,2992 H-K map, central 5\arcsec\,, pixels binned to 0\farcs1.
Contour levels are 2.6, 4.6, 6.6, 8.6  magnitudes of visual extinction A(V).
Images were smoothed with a 0\farcs45 gaussian before registering on the
nucleus and dividing.}
\label{n2extinction}
\end{figure}

The effects of dust absorption become noticeably less with increasing 
wavelength in our images of NGC\,2992. 
The K-band image has the most symmetrical appearance, with emission from
the galaxy disk visible furthest into the dust-lane.
By referencing fiducial colours of expected 
stellar populations to the actual observed
colours in NGC\,2992, we can estimate the extinction due to dust.
The H-K colour map is in general good for characterizing the extinction 
because it probes deeper within the dusty regions. However it has
the problem that K-band may be contaminated by hot dust (1000~K) emission,
resulting in possible overestimates  of the extinction. 
Our CO-narrowband imaging confirms previous speculation from ground-based
J-K and K-L colours (Alonso-Herrero \etal 1998) that hot dust is unlikely to
contribute significantly.

We assume that the colour of a typical early-type bulge
stellar population is $H-K=0.2$ (Glass \& Moorwood 1985), 
and that redder colours imply some degree of obscuration. Taking into account that the differential extinction between H and K is
$A(H) = 0.175 A(V)$ and
$A(K) = 0.112 A(V)$ (Rieke et al. 1985), we find
$$A(H-K) = - 2.5 \log(\frac{f(H)}{f(H_0)}) + 2.5 \log(\frac{f(K)}{f(K_0)})$$
where $f(K_0)/f(H_0)$ is the flux ratio corresponding to $H-K = 0.2$.
Thus $A(V) = A(H-K)/0.063$ in magnitudes.


Such an analysis is only true for extinction to the stars. Other emission 
processes (power law, emission line gas, etc.) will skew the results. 
The J-H and H-K colours show that there are two regions which can
likely not be explained simply by reddened stellar colours.
The first is within the dust lane, along the knotty spiral structure to the
south and to the north along the diffuse loop described in section 3.4.
The colours here tend to be bluer in both J-H and H-K, possibly indicative
of nebular emission.
The second is the extended region to the west of the nucleus where the colours are far from the locus of normal stellar colors, and different from both the 
disk/bulge to the east and the rest of the colours within the dust lane. 

The extinction map (figure 7) 
shows that if the central peak of emission is in fact due 
mainly to a compact stellar core, then there is an optical extinction of
A(V)$\sim$4 magnitudes. 
The J-H/H-K colours of the nucleus are consistent with
a reddened supergiant population. 
Taken at face value, the extinction map contours indicate A(V) extinction 
from 2.6 to 8.6 magnitudes with 2 magnitude intervals, although non-stellar
emission likely over-redden some of the circumnuclear regions as discussed above.
Using the J-H map results in an extinction estimate
which is lower, possibly as a result of the  larger optical depth at shorter
wavelengths.

\section{Discussion}

We have combined the various structures observed in the different 
wavelength regimes and 
depicted them schematically in figure 8. 

The radio loops lack any optical or near-IR counterparts except along the
galaxy disk axis. The bipolar outflows along the spin axis of the galaxy
observed at larger scales in the optical
(Allen \etal 1998) align along the same axis defined by the
radio loops. 
These facts suggest the actual figure-8 loops likely lie out of the galactic disk plane. 
However, the strong radio contours, emanating from the nucleus to the southwest,
lie along the buried spiral arm within the disk. 
Also, the red R-H loop to the north appears to be 
associated with the diffuse radio emission embedded within the larger radio loop.

Our hypothesis is then that the radio morphology consists of two components superimposed:\\
1) the loops out of the plane of the disk.\\
2) a component in the disk associated with the southern spiral arm,
 and a diffuse loop to the north. Starburst SNe remnants are the likely source
of these radio components.\\
The appearance of the radio figure-8 becomes more symmetrical if the galactic disk components (point 2 above) are subtracted, which supports such a superposition scenario.

The assumption of trailing large-scale spiral arms, 
in addition to the prominent dust lane likely lying in front of  the bulge, imply the NW edge of the galaxy disk 
is closer to us.
This scenario places the southern portion of the Extended Emission Line Region  (EELR) closer to us, with associated outflowing material (Allen \etal 1998). 
The southern radio loop would then also be closer to us, with the northern loop lying partially behind the dust lane.
However, the larger-scale spiral arms extending radially out past 30\arcsec\  
appear to wind in the opposite orientation to the inner spiral observed in our high resolution imagery.
This either forces the inner spiral arms to be leading, or else they are trailing in counter-rotation, with the inner region kinematically distinct from the outer galaxy.
The case is not clear from the velocity fields presented in Allen et al. (1998), which have low spatial resolution coupled with a complicated superposition of rotation and outflow components. 


%
%
\begin{figure}
\psfig{file=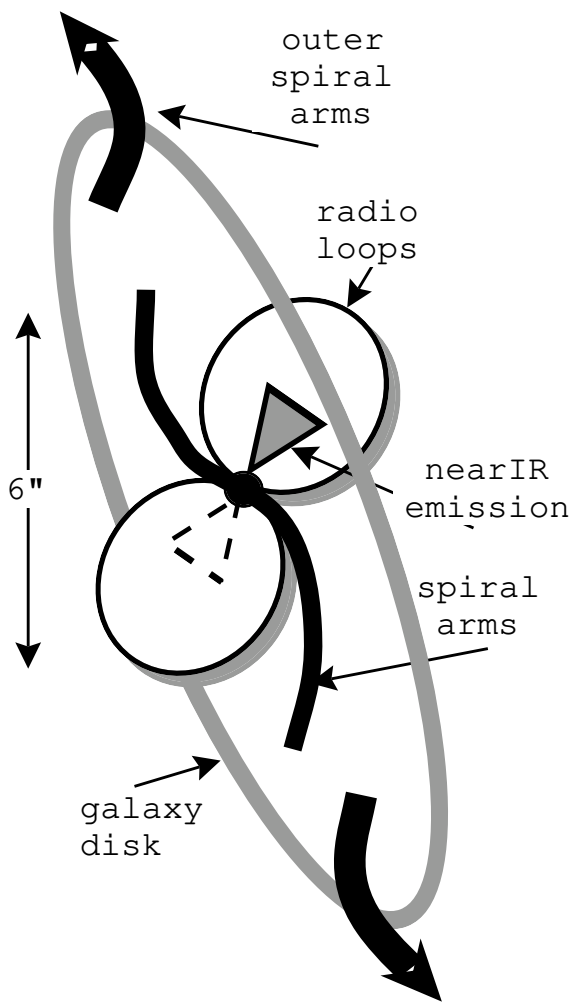,width=8.25cm,angle=0} 
\caption{NGC2992 cartoon displaying the hypothesized geometry in the central
7\arcsec ($\sim$ 1.3\,kpc). The spiral arms lie within the galaxy disk (large grey ellipse),
while the radio loops and possibly the nearIR extended emission 
are projected out of the plane of the disk. The speculative extended
emission inside the southern radio loop is depicted as a dashed cone.
The larger scale spiral arms extending out past 30\arcsec radially appear to wind in the
opposite orientation to the inner spiral. The assumption of trailing large-scale spiral arms, 
in addition the prominent dust lane likely lying in front of  the bulge, imply the 
NW edge of the disk is closer to us.
}
\label{fcartoon}
\end{figure}


The near-IR extended emission feature to the northwest gives the appearance of expanding into the northern radio loop, and the two emission features may be associated.
The morphology and the rather extreme near-IR colours suggest that an AGN driven jet, possibly with some continuum component, could be directed into the radio loop.
Hot dust present in an outflow is not ruled out for this feature either. 
This sort of near-IR ``jet" may exist towards the southern radio loop as well at a lower level.

\subsection{Interpretation}
As noted in the introduction, there have been several  models put forward for such figure-8 radio emission.
The most convincing in light of our new near-IR imaging is that the
figure-8 loops result from 
expanding  gas bubbles
which are seen preferentially as limb-brightened loops (Wehrle and Morris 1988).
The northern loop may be
related to the near-IR extended emission feature to the northwest.
Further evidence for the outflow picture exists in
the form of soft X-ray emission extended perpendicular to the disk of the 
galaxy Colbert \etal 1998), and H$\alpha$ imagery clearly showing the location 
of an extended emission line region (EELR) (Colbert \etal 1996).
These authors discuss two possible explanations
for the extended X-ray emission: 1) an AGN driven hot plasma, 2) a superwind from
a compact starburst. 

A galactic-scale superwind can be generated by either a compact starburst in
the galaxy core, or an AGN driven outflow which thermalizes the ISM at 
some distance from the core (Colbert et al. 1996).
In both cases the superwind would blow preferentially out of the galaxy 
plane where the pressure is lowest, such as observed in NGC253 (Unger et al. 1987).
With the resolved galaxy core in our images, it is not clear to what extent 
 dust absorption, hot dust emission, scattered AGN light,
or a compact stellar cluster contribute to the extended emission. 
The colours appear most consistent with reddened stellar light. 
Thus either picture could be consistent with our data since a stellar cluster 
and AGN (optical emission lines, flat radio spectrum)
are likely contributing to core emission processes.
However, the larger scale EELR need not be aligned with the 2\arcsec\ diameter
radio loops in the case of the AGN driven source.
The orientation of the EELR observed at larger scales (Allen et al. 1998) is in
fact roughly aligned with the radio loops.
For superwind models, the anisotropic EELR, seen in [OIII] and H$\alpha$, 
is likely to be associated with the outflow regions. However, the 
superwind in itself does
not yield a mechanism to produce continuum emission (Allen et al. 1998),
thus our near-IR data may rule out this latter model.


If the near-IR ``jet" is not actually related to the radio loops, 
the superwind model is still a plausible source of the loop emission.
The CO index provides evidence for a substantial population gradient in the 
core. We consider the case where the radio loops are  
due to a energetic burst of supernovae in the past. 
The luminosity of the stellar cluster must be at least three times that of 
the bubble so the shock can reach a galactic scale height
(Koo et al. 1992, Tenorio-Tagle et al. 1998)
The stellar cluster luminosity is estimated from the H-band image.
We determine the
size of the ``hypothetical" stellar cluster to be 0.5\arcsec\ with an absolute H-magnitude of $-16.5 + M_{AGN}$, where the AGN contribution is unknown and may be almost zero if the supermassive black hole is no longer being fueled, as described in the introduction (Bassini et al. 1998). The model PSF (section 3.1) 
scaled to the peak
of the extended core revealed that an unresolved AGN component could contribute at most 37\% to the emission within a 0.2\arcsec\ diameter aperture.


The second model outlined in the introduction, with the toroidal magnetic fields causing the
emission, is more difficult to explain in light of the near-IR extension and 
knotty emission along the southern radio loop.
The outflow driven bubble model explains the currently available data much
more naturally. In addition, a calculation of the magnetic energy in the loops
from the 8.4GHz VLA data (Falcke et al. 1997, Werhle et al. 1988), makes it difficult to model consistently in this
manner. Other problems with this model (Wehrle et al. 1988, Cameron 1985, Heyvaerts et al. 1987)
 associated with rotation timescales
and the lack of twisting of the radio/near-IR loops, make it even less
plausible.



\section{Conclusions}

We have presented adaptive optics near-IR and radio images of NGC2992 in conjunction
with archive HST optical imagery.
A spiral structure within the central 6\arcsec\ and a 1\arcsec\ 
extended feature are traced down to the core at the resolution of our images.
We speculate that multiple radio components are superposed which contribute to the
observed figure-8 morphology in the VLA images: one associated with the
spiral structure in the galaxy disk, and another flowing out of the
galaxy plane.

IR and optical spectra at high spatial resolution will likely 
provide the means of determining if the
population gradients in the core of NGC\,2992 are due to changes
in age and/or metallicity.
Such spectral imagery will also permit the nature of the extended 
structures to be explored, shedding light on the possible connection to 
the radio loops.

Our current hypothesis concerning the radio loops  involves an AGN
outflow powering the loop rather than a starburst superwind, as any
near-IR emission related to the jet 
would be unlikely in the latter case.
NGC\,2992 represents yet another example of star formation and AGN components
both existing in the galaxy core (Storchi-Bergman \etal 1997). There is no
obvious indication in our data 
of whether there is any connection between the two in
evolutionary terms.

\subsection*{ACKNOWLEDGEMENTS}

We would like to acknowledge the staff at CFHT and VLA for facilitating
these observations. The CADC database was invaluable in obtaining HST images.

\end{document}